\documentclass[11pt,a4paper]{article}
\def\bea{\begin{eqnarray}}
\def\eea{\end{eqnarray}}
\def\nn{\nonumber}
\def\beq{\begin{equation}}
\def\eeq{\end{equation}}
\def\ba{\beq\new\begin{array}{c}}
\def\ea{\end{array}\eeq}
\def\be{\ba}
\def\ee{\ea}

\makeatletter
\newdimen\normalarrayskip              
\newdimen\minarrayskip                 
\normalarrayskip\baselineskip
\minarrayskip\jot
\newif\ifold             \oldtrue            \def\new{\oldfalse}
\def\arraymode{\ifold\relax\else\displaystyle\fi} 
\def\eqnumphantom{\phantom{(\theequation)}}     
\def\@arrayskip{\ifold\baselineskip\z@\lineskip\z@
     \else
     \baselineskip\minarrayskip\lineskip2\minarrayskip\fi}
\def\@arrayclassz{\ifcase \@lastchclass \@acolampacol \or
\@ampacol \or \or \or \@addamp \or
   \@acolampacol \or \@firstampfalse \@acol \fi
\edef\@preamble{\@preamble
  \ifcase \@chnum
     \hfil$\relax\arraymode\@sharp$\hfil
     \or $\relax\arraymode\@sharp$\hfil
     \or \hfil$\relax\arraymode\@sharp$\fi}}
\def\@array[#1]#2{\setbox\@arstrutbox=\hbox{\vrule
     height\arraystretch \ht\strutbox
     depth\arraystretch \dp\strutbox
     width\z@}\@mkpream{#2}\edef\@preamble{\halign
\noexpand\@halignto
\bgroup \tabskip\z@ \@arstrut \@preamble \tabskip\z@ \cr}%
\let\@startpbox\@@startpbox \let\@endpbox\@@endpbox
  \if #1t\vtop \else \if#1b\vbox \else \vcenter \fi\fi
  \bgroup \let\par\relax
  \let\@sharp##\let\protect\relax
  \@arrayskip\@preamble}
\def\eqnarray{\stepcounter{equation}%
              \let\@currentlabel=\theequation
              \global\@eqnswtrue
              \global\@eqcnt\z@
              \tabskip\@centering
              \let\\=\@eqncr
              $$%
 \halign to \displaywidth\bgroup
    \eqnumphantom\@eqnsel\hskip\@centering
    $\displaystyle \tabskip\z@ {##}$%
    \global\@eqcnt\@ne \hskip 2\arraycolsep
         $\displaystyle\arraymode{##}$\hfil
    \global\@eqcnt\tw@ \hskip 2\arraycolsep
         $\displaystyle\tabskip\z@{##}$\hfil
         \tabskip\@centering
    &{##}\tabskip\z@\cr}
\begingroup\ifx\undefined\newsymbol \else\def\input#1 {\endgroup}\fi
\input amssym.def \relax
\input amssym
\newfont{\hr}{msbm10}
\newfont{\ams}{msam10}

\font\numbers=cmss12
\font\upright=cmu10 scaled\magstep1
\def\stroke{\vrule height8pt width0.4pt depth-0.1pt}
\def\topfleck{\vrule height8pt width0.5pt depth-5.9pt}
\def\botfleck{\vrule height2pt width0.5pt depth0.1pt}
\def\Zmath{\vcenter{\hbox{\numbers\rlap{\rlap{Z}\kern 0.8pt\topfleck}\kern 2.2pt
                   \rlap Z\kern 6pt\botfleck\kern 1pt}}}
\def\Qmath{\vcenter{\hbox{\upright\rlap{\rlap{Q}\kern
                   3.8pt\stroke}\phantom{Q}}}}
\def\Nmath{\vcenter{\hbox{\upright\rlap{I}\kern 1.7pt N}}}
\def\Cmath{\vcenter{\hbox{\upright\rlap{\rlap{C}\kern
                   3.8pt\stroke}\phantom{C}}}}
\def\Rmath{\vcenter{\hbox{\upright\rlap{I}\kern 1.7pt R}}}
\def\Z{\ifmmode\Zmath\else$\Zmath$\fi}
\def\Q{\ifmmode\Qmath\else$\Qmath$\fi}
\def\N{\ifmmode\Nmath\else$\Nmath$\fi}
\def\C{\ifmmode\Cmath\else$\Cmath$\fi}
\def\R{\ifmmode\Rmath\else$\Rmath$\fi}

\newcounter{app}

\def\app{\setcounter{equation}{0}
\def\theequation{\Alph{app}.\arabic{equation}}\par
   \addvspace{4ex}
   \@afterindentfalse
  \secdef\@app\@dapp}

\newcommand\@app{\@startsection {app}{1}{0ex}%
                                   {-3.5ex \@plus -1ex \@minus -.2ex}%
                                   {2.3ex \@plus.2ex}%
                                   {\normalfont\Large\bf}}
\def\@dapp#1{%
{\parindent \z@ \raggedright  \bf #1}\par\nobreak}
\def\l@app#1#2{\ifnum \c@tocdepth >\z@
    \addpenalty\@secpenalty
    \addvspace{1.0em \@plus\p@}%
    \setlength\@tempdima{8em}%
    \begingroup
      \parindent \z@ \rightskip \@pnumwidth
      \parfillskip -\@pnumwidth
      \leavevmode \bfseries
      \advance\leftskip\@tempdima
      \hskip -\leftskip
      #1\nobreak\hfil \nobreak\hb@xt@\@pnumwidth{\hss #2}\par
    \endgroup\fi}
\newcounter{sapp}[app]

\def\sapp{\def\theequation{\Alph{app}.\arabic{equation}}
\par
\@afterindentfalse
  \secdef\@sapp\@dsapp}
\newcommand{\@sapp}{\@startsection{sapp}{2}{\z@}%
                                     {-3.25ex\@plus -1ex \@minus 
-.2ex}%
                                     {1.5ex \@plus .2ex}%
                                     {\normalfont\large\bfseries}}

\def\@dsapp#1{%
{\parindent \z@ \raggedright  \bf #1
}\par\nobreak}
\newcommand{\l@sapp}{\@dottedtocline{2}{1.5em}{2.3em}}

\def\2{{1\over 2}}
\def\N2{${\cal N}=2$}

\def\be{ \begin{eqnarray} }
\def\ee{ \end{eqnarray} }

\def\bea{\begin{eqnarray}}
\def\eea{\end{eqnarray}}
\def\nn{\nonumber}

\def\beq{\begin{equation}}
\def\eeq{\end{equation}}
\def\ba{\beq\new\begin{array}{c}}
\def\ea{\end{array}\eeq}
\def\be{\ba}
\def\ee{\ea}

\title{
Experiments with the WDVV equations 
\\
for the gluino-condensate prepotential:
\\
the cubic (two-cut) case
}

\author{
H.Itoyama$^{1}$ and A.Morozov$^{2}$ \
\\ \normalsize \em $^{1}$ 
Department of Mathematics and Physics, Osaka City University, Osaka, Japan
\\
\normalsize \em $^{2}$
Institute of Theoretical and Experimental Physics, Moscow}

\date{November, 2002}

\begin{document}

\maketitle

\vspace{-9.7cm}

\begin{center}
\hfill hep-th/0211259 \\
\hfill OCU-PHYS 195\\
\hfill ITEP/TH-58/02
\end{center}

\vspace{6.5cm}

\begin{abstract}
We demonstrate by explicit calculation
that the first two terms in the CIV-DV
prepotential for the two-cut case satisfy the
generalized WDVV equations, just as in all other known
examples of hyperelliptic Seiberg-Witten models.
The WDVV equations are non-trivial in this situation,
provided the set of moduli is extended as compared to
the Dijkgraaf-Vafa suggestion and includes also
moduli, associated with the positions of the cuts
(not only with their lengths). Expression for
the extra modulus dictated by WDVV equation, however,
appears different from a naive expectation implied by
the Whitham theory. Moreover, for every value of the
"quantum-deformation parameter" $1/g_3$,
we actually find an entire one-parameter family of
solutions to the WDVV equations, of which the
conventional prepotential is just a single point.
\end{abstract}

\vspace{2.0cm}


In this paper we report on some "experimental" results,
confirming the validity of the WDVV equations 
for the realistic prepotential of gluino-condensate fields 
$S_i$ \cite{CIV}. 
We follow the program suggested in our recent paper
\cite{IM4}, which implies, among other things,  
that the considerations of
R.Dijkgraaf and C.Vafa \cite{DV}
(see \cite{DVfollowup} for the further developements) should be 
supplemented to include extra moduli, associated with
positions of the cuts in the planar matrix model solution.
The purpose of the present paper is to confirm, that
the "spherical" WDVV equations \cite{WDVV,MMM,WDVVfollowup} 
are indeed satisfied after these moduli are included, as expected
on general grounds in all hyperelliptic examples of
Seiberg-Witten theory \cite{SW} and its interpretation
in terms of integrable systems \cite{GKMMM}.
However, it turns out that the naive expectation of \cite{IM4}
for explicit choice of these extra $T$-moduli should be
modified already at the level of the two-cut solution
(when there are just two $S$-moduli and one extra $T$-modulus).
Interpretation of these results and their generalization
to higher number of cuts $n$ remain for the subject of further analysis.
 
\section{Introduction}

The prepotential, which will be the subject of our
discussion,  has the form \cite{CIV,DV}:
 
\be
{\cal F}(S,T) = {\cal F}_{pert}(S,T) + 
\sum_{k=1}^\infty g_{n+1}^{-k}{\cal F}_{k+2}(S,T)
\ee
where ${\cal F}_{k+2}(S,T)$ is a polynomial of degree $k+2$ in $S$'s,
with coefficients made from $\alpha_{ij} = \alpha_i-\alpha_j$ 
(see (\ref{Ppol}) below) and these $\alpha_{ij}$ are somehow incorporated
into the $n-1$ auxiliary {\it flat} moduli $T_i$.
It can be obtained after throwing away some unwanted terms from the
Dijkgraaf-Vafa prepotential, built by the common rules of the SW
theory from a family of hyperelliptic complex curves (Riemann surfaces)

\be
Y^2 = P_n^2(x) + P_n(x)\sum_{i=1}^n\frac{\tilde S_i(x)}{x-\alpha_i},
\ \ P_n(x) = \prod_{i=1}^n (x-\alpha_i)
\label{Ppol}
\ee
with the one form

\be
dS_{DV} = Y(x)dx
\ee
defined on every curve. The {\it flat}
moduli $S_i$, $i=1,\ldots,n$ are infinite series in integer
powers of $\tilde S_i$ with $\alpha$-dependent coefficients,
while $\alpha$-dependence is somehow expressed through the
remaining $n-1$ {\it flat} moduli $T_i$, $i=1,\ldots,n-1$.
We refer to \cite{IM4} for further details.  

The perturbative (logarithmic) piece of the prepotential can
be easily obtained from

\be
 \frac{\partial{\cal F}_{pert}}{\partial S_i} =
  \int_{\gamma_i+\rho_i}^\Lambda dS_{pert} =  \cdots   +
S_i\log\frac{\Lambda}{\rho_i} +
\sum_{j\neq i} S_j\log\frac{\Lambda}{\alpha_{ij}}
\label{pertprep}
\ee
with

\be
dS_{pert}(x) = P_n(x) dx +  \sum_{i=1}^n \frac{S_i}{x-\alpha_i}dx
\ee
and

\be
 ( i \pi g_{n+1})\rho_i^2 \prod_{j\neq i}^n\alpha_{ij} = S_i 
+ O(S^2)
\label{rhothroughS}
\ee
(eq.(\ref{rhothroughS}) is the condition that $\alpha_i\pm\rho_i$
approximate -- in the leading order in small $S$'s -- the
ramification points of the curve (\ref{Ppol})).
Altogether, these formulas imply that\footnote{
The terms $ g_{n+1} \left( W(\Lambda)\sum_i S_i 
 -\sum_i  W(\alpha_{i}) S_i \right) $
with $W'(x) = P_n(x)$ and 
$(\sum_{i} S_i)^2\log\Lambda$ are omitted from this expression.
These are harmless for considerations below as they are at most
quadratic in moduli and drop out of the prepotential's third 
derivatives. Note that $\tilde {\tilde T_k} =  W(\alpha_{i}) $ 
 can be an interesting possibility  to try for the role of
 $T$ moduli.
}

\be
2\pi i  {\cal F}_{pert}(S,T) =
\frac{1}{2} \sum_{i=1}^n S_i^2\log S_i -
\frac{1}{2}\sum_{i<j}^n(S_i^2 - 4S_iS_j + S_j^2)\log \alpha_{ij}
\label{perpre}
\ee

In what follows we present some experimental evidence that
the WDVV equations in the form, suggested in ref.\cite{MMM},

\be
\check{\cal F}_I \check{\cal F}_J^{-1} \check{\cal F}_K
= \check{\cal F}_K \check{\cal F}_J^{-1} \check{\cal F}_I
\ee
(with matrices $\check{\cal F}_L$ made out of the third
derivatives of the prepotential ${\cal F}(a_I)$,
$\left(\check{\cal F}_L\right)_{MN} = {\cal F}_{LMN}
= \partial^3{\cal F}/\partial a_L\partial a_M\partial a_N$),
can still hold for this prepotential, at least
for the cubic superpotential, when there are just two $S$-moduli
and one $T$-modulus.  Dependence on the
$T$-modulus which is not-quite-expected is needed, however,
 to make these equations true. 
Instead of being equal to $\tilde T_1 \sim \alpha_1\alpha_2 =
-\frac{1}{2}\alpha_{12}^2$, according to the general rule
\cite{IM2,GMMM,IM4}\footnote{
From experience with the other SW models \cite{GMMM} one can 
conjecture that the $x^{-n}$ factors in (\ref{T-def})
can be in fact substituted by $w^{-k/n}$
(see \cite{IM4} for the definition of $w$), but this difference
is inessential for our considerations in this paper.
}

\be
\tilde T_k = \frac{1}{k}res_\infty x^{-k}dS(x) = 
\frac{1}{k}res_\infty x^{-k}P_n(x)dx,
\label{T-def}
\ee
it appears (experimentally!) that $T\sim \alpha_{12}^3$. (Note that
another naive choice $T_i = \alpha_i-\alpha_n$ is also ruled out).
It remains for the future research to find an interpretation of this
metamorphosis and, even more important, to understand if it can be
extended to higher corrections and more sophisticated superpotentials
(with $n>2$). For $n>2$ it is not easy even to reexpress
the differences $\alpha_{ij}$ through
$n-1$ {\it flat} moduli $T_i$, which are expected to be polynomial
 in $\alpha_{ij}$.

From now on we consider just the case of $n=2$, but introduce some
free parameters into (\ref{perpre}) and into ${\cal F}_3$ in order
to find what kind of restrictions are imposed on these parameters
by the WDVV equations.

\section{Perturbative $n=2$ case}

Let

\be
{\cal F}_{pert}(S_1,S_2,T) =
\frac{1}{2}S_1^2\log S_1 + \frac{1}{2}S_2^2\log S_2 +
\frac{\nu}{2}(S_1^2 - 2bS_1S_2 + S_2^2)\log T
\label{}
\ee
(actually, in DV theory  $b=2$ and $T^{-\nu}=\alpha_{12}$).
Then

\be
\frac{\partial {\cal F}_{pert}}{\partial S_1} =
S_1\log S_1 + \nu(S_1-bS_2)\log T,\nn \\
\frac{\partial {\cal F}_{pert}}{\partial S_2} =
S_2\log S_2 + \nu(S_2-bS_1)\log T,\nn \\
\frac{\partial {\cal F}_{pert}}{\partial T} =
\frac{\nu}{2T}(S_1^2 - 2bS_1S_2 + S_2^2),
\ee
and matrices (we omit the index {\it perp} in what follows)
\be
\check{\cal F}_1 =
\left(
\begin{array}{ccc}
\frac{1}{S_1} & 0 &
\frac{\nu}{T}
\\
0 & 0 & -\frac{\nu b}{T}
\\
\frac{\nu}{T} & -\frac{\nu b}{T} & -\frac{\nu (S_1-bS_2)}{T^2}
\end{array}
\right)
\ee

\be
\check{\cal F}_2 =
\left(
\begin{array}{ccc}
0 & 0 & -\frac{\nu b}{T} 
\\
0 & \frac{1}{S_2} & \frac{\nu}{T}
\\
-\frac{\nu b}{T} & \frac{\nu}{T} & -\frac{\nu (S_2-bS_1)}{T^2}
\end{array}
\right)
\ee

\be
\check{\cal F}_T =
\left(
\begin{array}{ccc}
\frac{\nu}{T} & -\frac{\nu b}{T} & -\frac{\nu (S_1-bS_2)}{T^2}
\\
-\frac{\nu b}{T} & \frac{\nu}{T} & -\frac{\nu (S_2-bS_1)}{T^2}
\\
-\frac{\nu (S_1-bS_2)}{T^2} & -\frac{\nu (S_2-bS_1)}{T^2} &
\frac{\nu (S_1^2 -2bS_1S_2 + S_2^2)}{T^3}
\end{array}
\right)
\ee
Further,

\be
\check{\cal F}_2^{-1} = 
\left(
-\frac{\nu^2 b^2}{S_2 T^2}
\right)^{-1}
\left(
\begin{array}{ccc}
-\frac{\nu(\nu + 1)}{T^2} + \frac{\nu bS_1}{S_2T^2} &
-\frac{\nu^2b}{T^2} & \frac{\nu b}{S_2 T}
\\
-\frac{\nu^2b}{T^2} & -\frac{\nu^2b^2}{T^2} & 0
\\
\frac{\nu b}{S_2 T} & 0 & 0
\end{array}
\right);
\ee
and
 
\be
\check{\cal F}_1\check{\cal F}_2^{-1} = 
\left(
-\frac{\nu^2 b^2}{S_2 T^2}
\right)^{-1}
\left(
\begin{array}{ccc}
-\frac{\nu(\nu + 1)(S_2-bS_1)}{S_1S_2T^2} &
-\frac{\nu^2b}{S_1T^2} & \frac{\nu b}{S_1S_2 T}
\\
-\frac{\nu^2b^2}{S_2T^2} & 0 & 0
\\
\frac{\nu^2(\nu + 1)(b^2-1)}{T^3} &
\frac{\nu^3 b(b^2-1)}{T^3} & 
\frac{\nu^2b}{S_2T^2}
\end{array}
\right) 
= \nn \\ =
\left(
\begin{array}{ccc}
\frac{(\nu + 1)(S_2-bS_1)}{\nu b^2S_1} &
\frac{S_2}{bS_1} & -\frac{T}{\nu bS_1}
\\
1 & 0 & 0
\\
-\frac{(\nu + 1)(b^2-1)S_2}{b^2T} &
-\frac{\nu (b^2-1)S_2}{bT} & 
-\frac{1}{b}
\end{array}
\right),
\label{F1/F2}
\ee

\be
\check{\cal F}_2^{-1}\check{\cal F}_T = 
\left(
\begin{array}{ccc}
-\frac{(\nu + 1)(b^2-1)S_2}{b^2T} & 0 &
\frac{(\nu + 1)(b^2-1)S_1S_2}{b^2T^2} 
\\
-\frac{\nu (b^2-1)S_2}{bT} & 0 &
\frac{\nu (b^2-1)S_1S_2}{bT^2} 
\\
-\frac{1}{b} & 1 & \frac{S_1-bS_2}{bT}
\end{array}
\right)
\label{FT/F2}
\ee
Finally,

\be
\check{\cal F}_1\check{\cal F}_2^{-1} \check{\cal F}_T =
\left(
\begin{array}{ccc}
* & \frac{\nu}{T} & 
\frac{\nu S_1}{bT^2} + \xi\frac{S_2}{b^2T^2}
\\
\frac{\nu}{T} & * & -\frac{\nu (S_1 - bS_2)}{T^2}
\\
\frac{\nu S_1}{bT^2} + \eta\frac{S_2}{b^2T^2}
& -\frac{\nu (S_1 - bS_2)}{T^2} & *
\end{array}
\right)
\label{triple}
\ee
with

\be
\xi = b^2-1-\nu, \nn \\
\eta = \nu^2(b^2-1)^2-\nu(2b^2-1).
\ee
Stars stand for diagonal elements that we do not need.\footnote{
Zeroes in the second row of (\ref{F1/F2}) and 
the second column in (\ref{FT/F2}) are pertinent
properties of all matrices $\check{\cal F}$ composed of third derivatives,
they are independent of particular ansatz for ${\cal F}$. The same
is true with coincidence of entries $12$ and $21$,
$23$ and $32$ in (\ref{triple}): the only non-trivial condition
  obtained from (\ref{triple}) being symmetric
(WDVV equation), imposed non-trivially on ${\cal F}$ 
in the $n=3$ case, is an equality between the entries $13$ and $31$.
}
This matrix is symmetric, provided $b$ and $\nu$ are related by the
equation $\xi = \eta$:

\be
\nu^2(b^2-1)^2-\nu(2b^2-1) = (b^2-1-\nu),
\ee
or

\be
\nu^2(b^2-1)-2\nu - 1 = 0,
\label{nubrelation}
\ee
or 

\be
\nu = \pm\frac{1}{b\mp 1}
\ee
For $b=2$ this gives two solutions: $\nu = 1$ and $\nu = -\frac{1}{3}$.
The second choice, $\nu = -\frac{1}{3}$, is nicely consistent with
the homogeneity of the prepotential, but it implies that $T=\alpha_{12}^3$,
what seems inconsistent with any {\it a priori} formula like (\ref{T-def}).

\section{Next order: $n=2$ case}

For corrected prepotential

\be
{\cal F}_{pert}(S_1,S_2,T) + f(S_1,S_2,T)
\label{firstcor}
\ee
we need that

\be
{\cal A}=
\check f_1 \check{\cal F}_2^{-1} \check{\cal F}_T +
\check{\cal F}_1\check{\cal F}_2^{-1} \check f_T  -
\check{\cal F}_1\check{\cal F}_2^{-1} \check f_2 
\check{\cal F}_2^{-1} \check{\cal F}_T 
+ O(f^2)
\label{nocorr}
\ee
 be symmetric.

Take the first "non-perturbative" correction in the form

\be
f(S_1,S_2,T) \sim \frac{1}{g_3}{\cal F}_3(S_1,S_2,T) \sim
(S_1^3 - cS_1^2S_2 + cS_1S_2^2 - S_2^3)T^{3\mu}
\label{fcorans}
\ee
The overall coefficient like $g_3^{-1}$ does not play a role in
the linearized equation  (\ref{nocorr}). The ansatz is consistent with
\cite{CIV}, according to their calculation 
$c=-\frac{15}{2}$ and $T^{-\mu}=\alpha_{12}$. (Since at the perturbative
level we already obtained $T=\alpha_{12}^3$ for the CIV-DV prepotential.
This implies that we need $\mu = -1/3$ for it). 
Substituting the ansatz (\ref{fcorans}) into (\ref{nocorr}) and
making use of (\ref{F1/F2}) and (\ref{FT/F2}), we obtain

\be
{\cal A} =
\left(\begin{array}{ccc}
f_{111} & f_{112} & f_{11T}
\\
f_{122} & f_{222} & f_{12T}
\\
f_{11T} & f_{12T} & f_{1TT}
\end{array}\right)
\left(
\begin{array}{ccc}
-\frac{(\nu + 1)(b^2-1)S_2}{b^2T} & 0 &
\frac{(\nu + 1)(b^2-1)S_1S_2}{b^2T^2} 
\\
-\frac{\nu (b^2-1)S_2}{bT} & 0 &
\frac{\nu (b^2-1)S_1S_2}{bT^2} 
\\
-\frac{1}{b} & 1 & \frac{S_1-bS_2}{bT}
\end{array}
\right) +
\nn \\ +
\left(
\begin{array}{ccc}
\frac{(\nu + 1)(S_2-bS_1)}{\nu b^2S_1} &
\frac{S_2}{bS_1} & -\frac{T}{\nu bS_1}
\\
1 & 0 & 0
\\
-\frac{(\nu + 1)(b^2-1)S_2}{b^2T} &
-\frac{\nu (b^2-1)S_2}{bT} & 
-\frac{1}{b}
\end{array}\right)\cdot
\label{prom}
\\ \cdot
\left(
\begin{array}{ccc}
\frac{(b^2-1)S_2}{b^2T}((\nu+1)f_{112} +b\nu f_{122}) +
& 0 &
-\frac{(b^2-1)S_1S_2}{b^2T^2}((\nu+1)f_{112} +b\nu f_{122}) + 
\\
+ \frac{1}{2}f_{12T} + f_{11T} & & 
+ \frac{2S_2-S_1}{2T}f_{12T} + f_{1TT}
\\ & & \\
\frac{(b^2-1)S_2}{b^2T}((\nu+1)f_{122} +b\nu f_{222}) +
& 0 &
-\frac{(b^2-1)S_1S_2}{b^2T^2}
((\nu+1)f_{122} +b\nu f_{222}) +
\\
+\frac{1}{2}f_{22T} + f_{12T} & &
+ \frac{2S_2-S_1}{2T}f_{22T} + f_{2TT}
\\ & & \\
\frac{(\nu + 1)(b^2-1)S_2}{b^2T} f_{12T}
+ \frac{\nu (b^2-1)S_2}{bT} f_{22T} +
& 0 &
-\frac{(\nu + 1)(b^2-1)S_1S_2}{b^2T^2} f_{12T}
- \frac{\nu (b^2-1)S_1S_2}{bT^2} f_{22T} + 
\\
+ \frac{1}{2}f_{2TT} + f_{1TT} & &
+ \frac{2S_2-S_1}{2T}f_{2TT} + f_{TTT}
\end{array}\right) 
\nn
\ee

The last matrix at the r.h.s. is obtained by evaluating the sum

\be
\left\{
\left(\begin{array}{ccc}
f_{11T}
&  f_{12T}
&  f_{1TT}
\\
f_{12T}
& f_{22T}
& f_{2TT}
\\
f_{1TT}
& f_{2TT}
& f_{TTT}
\end{array}\right) -
\right. \nn \\ \left.
-
\left(\begin{array}{ccc}
f_{112} & f_{122}   & f_{12T}
\\
f_{122} & f_{222} & f_{22T}
\\
f_{12T} & f_{22T} & f_{2TT}
\end{array}\right)
\left(
\begin{array}{ccc}
-\frac{(\nu + 1)(b^2-1)S_2}{b^2T} & 0 &
\frac{(\nu + 1)(b^2-1)S_1S_2}{b^2T^2} 
\\
-\frac{\nu (b^2-1)S_2}{bT} & 0 &
\frac{\nu (b^2-1)S_1S_2}{bT^2} 
\\
-\frac{1}{b} & 1 & \frac{S_1-bS_2}{bT}
\end{array}
\right)
\right\}
\ee

That  matrix (\ref{prom}) be symmetric can seem to impose
three relations. In fact, two of them
${\cal A}_{12} = {\cal A}_{21}$ and
${\cal A}_{23} = {\cal A}_{32}$ are automatically satisfied
(for the composition of $3\times 3$ matrices made from third
derivatives, actually ${\cal A}_{12} = f_{11T}$ and 
${\cal A}_{23} = f_{1TT}$).
The only non-trivial one is the equality between the
$13$ and $31$ entries, ${\cal A}_{13} - {\cal A}_{31} = 0$.
After a tedious calculation we obtain for this difference
with the ansatz (\ref{fcorans}):

\be
b^2S_1T^{3-3\mu}({\cal A}_{13} - {\cal A}_{31}) =
\nn \\ =
\left[bS_1^2 +
(2(b^2-1)\nu + (b^2-2))S_1S_2
\right] \cdot 
6\mu(3S_1-cS_2)
+ \nn \\ 
+ \left[
\frac{(\nu + 1)}{\nu} (S_1^2 + S_2^2) 
+ \frac{2
((b^4-1)\nu^2 + (b^2-2)\nu -1)}{\nu b}
S_1S_2  
\right] \cdot 
6\mu c(S_2-S_1)
+ \nn \\ 
\left[bS_2^2 +
(2(b^2-1)\nu + (b^2-2))S_1S_2)
\right] \cdot 
6\mu(cS_1-3S_2)
+ \nn \\ 
+ \left[
\frac{\nu +1}{\nu}S_2 + \frac{\nu -1}{\nu}bS_1
\right] \cdot 
3\mu(3\mu -1)(3S_1^2 - 2cS_1S_2 + cS_2^2)
+ \nn \\ 
+ \left[
\frac{\nu -1}{\nu}bS_2 + \frac{\nu +1}{\nu}S_1
\right] \cdot
3\mu(3\mu -1)(-cS_1^2 + 2cS_1S_2 - 3S_2^2) 
+ \nn \\
+ 3b\mu(3\mu -1)(3\mu -2)
(-S_1^3 + cS_1^2S_2 - cS_1S_2^2 + S_2^3)
+ \nn \\
+ \left[
\frac{6(\nu+1)(b^2-1)}{b^2} -
\frac{2(b^2-1)[(b^2-b+1)\nu^2-(b-2)\nu+1]}{b\nu}c
\right] \cdot 
S_1S_2
+ \nn \\ 
+ (b^2-1)((b^2-1)\nu-1)
\left[
-6\nu + \frac{2(\nu+1)[(2b-1)\nu-1]}{\nu b^2}c
\right] \cdot 
S_2^2
= \nn \\ \nn \\ \nn \\ =
(3\mu +1)\left\{\frac{3\mu}{\nu}[(\nu+1)c - b(3\nu-3\mu+1)](S_2^3-S_1^3)
+ \right.\nn\\ \left. +
\left(\left[3\mu(3\mu-1)b^2+2(3\mu-1)b + 4 +
\right.\right.\right.\nn\\ \left.\left.\left.
+(2(b^3-3b+4) - 3(3b-2)b\mu)\nu) -
4(b^2-1)(b^2-b+1)\nu^2\right]c 
+ \right.\right.\nn\\ \left.\left. +
[3\mu(\nu +1) + 4(b^2-1)\nu^2 + 2(b^2-3)\nu -2]
\right)S_1^2S_2
+ \right.\nn\\ \left. +
\left(\left[\frac{4((b^4-1)\nu^2+(b^2-2)\nu-1)}{\nu b}
-\frac{(3\mu-2)^2b}{\nu} - 4b^2v + 4\nu -
\right.\right.\right.\nn\\ \left.\left.\left.
- 2b^2 + 2
-\frac{2}{\nu} + 2b + (3\mu-2)\frac{3b(\nu-1)-2(\nu+1)}{\nu}
\right]c
- \right.\right.\nn\\ \left.\left. 
\phantom{\frac{3\mu}{2}}
-\ [12(b^2-1)\nu + 6(b^2-2) + (\nu+1)(3\mu-2)]
\right)S_1S_2^2
\right\}
+ \nn \\  \nn \\ +
((b^2-1)\nu^2-2\nu-1)\left\{
\left(\frac{b^2-b+1}{b\nu}c - \frac{6}{\nu}\right)S_1^2S_2
+ \right.\nn\\ \left. +
\left(\frac{2(2b-1)}{b^2\nu}c - \frac{6}{\nu}\right)((b^2-1)\nu-1)S_1S_2^2
\right\}
\label{fcorrwdvv}
\ee
The r.h.s. is a linear combination of sophisticated expressions\footnote{
In fact, the overloaded formulas at the r.h.s. of
(\ref{fcorrwdvv}) can still contain errors
inside the curly brackets; what we were
careful to check is the fact that the coefficients are indeed equal to
$3\mu+1$ and $(b^2-1)\nu^2-2\nu-1$, what is the only thing important
for our discussion. Still, the formulas deserve independent check
with the help of Mathmatica or Maple.
}
with coefficients equal to $3\mu+1$ and $(b^2-1)\nu^2-2\nu-1$.
Thus it vanishes, and WDVV eqs hold in this approximation
whenever $3\mu = -1$ and the condition (\ref{nubrelation})
for logarithmic WDVV to hold is fulfilled. No restriction
on parameter $c$ appears in this approximation. 

\section{Conclusion}

To conclude, we proved that perturbative prepotentials of the
given type, satisfying the {\it spherical} WDVV equations
of ref.\cite{MMM}, do exist, 
at least in the case of just two $S$-moduli and one $T$-modulus
and at the first order of expansion in powers of $S$. 

Moreover, solutions form an entire one-parametric family
(\ref{nubrelation}), of which the CIV-DV prepotential is just
a single point $b=2$. It is a question, what kind of SW 
system can correspond to generic solution with any $b$.

Furthermore, we demonstrated that,
as usual with the WDVV equations, once they are satisfied at
the perturbative (logarithmic) level, the prepotential can be corrected
so that they are satisfied at the next order -- and, perhaps, further
and further, so that every logatithmic solution of WDVV eqs gives rise
to (at least a one-parametric, with dimensionful parameter $g_{n+1}$) 
family of solutions. Actually, to the approximation that we considered,
one more dimensionless parameter, $c$, remained unrestricted, 
but it can still be fixed in the next approximations.

All these imply that the DV system does not provide a {\it generic}
solution of the given type to WDVV equations.
One can regard this as a confirmation of the universality-classes
hypothesis of ref.\cite{IM4} (see the end of s.5.1 of that paper).
This emphasizes the need to study {\it regularized} DV system,
the "hidden-sector" curve and the DV limiting procedure (see
\cite{IM4} for more details).

\section{Acknowledgements}

A.M. acknowledges the support of JSPS and the hospitality
of the Osaka City University during his stay at Japan.
Our work is partly supported by the 
Grant-in-Aid for Scientific Research (14540284) from the
Ministry of Education, Science and Culture, Japan (H.I),
and by the Russian President's grant 00-15-99296, RFBR-01-02-17488,
INTAS 00-561 and by Volkswagen-Stiftung (A.M.).

\end{document}